# Temperature and terahertz frequency dependence of the dielectric properties of $Fe_3O_4$ thin films deposited on Si substrate

Ashish Khandelwal[1,*], L. S. Sharath Chandra[1], Shilpam Sharma[1], Archna Sagdeo[2,5], J. Jayabalan[3,5], R. J. Choudhary[4] and M. K. Chattopadhyay[1, 5]

[1]*Free Electron Laser Utilization Laboratory, Raja Ramanna Centre for Advanced Technology, Indore 452 013, India.*

[2]*Synchrotrons Utilization Section, Raja Ramanna Centre for Advanced Technology, Indore 452 013, India*

[3]*Nano Science Laboratory, Material Science Section, Raja Ramanna Centre for Advanced Technology, Indore 452 013, India.*

[4]*UGC–DAE Consortium for Scientific Research, Khandwa Road, University Campus, Indore 452 001, India.*

[5]*Homi Bhabha National Institute, Training School Complex, Anushaktinagar, Mumbai 400 094, India.*

* *ashishkhandelwal@rrcat.gov.in*

## Abstract

The $Fe_3O_4$/Si films are considered to be promising materials for THz spintronic applications due to their high temperature magnetic transition and semiconducting properties. In this article, we present the real part of the dielectric constant ($\varepsilon_1$) and the optical conductivity ($\sigma_1$) of $Fe_3O_4$ films of different thicknesses deposited on Si substrate ($Fe_3O_4$/Si) in the THz range at temperatures 2- 300 K. Although the magnetization of the films with thickness ≥ 115 nm shows a clear change at the Verwey transition temperature $T_v$ = 121 K, their optical properties in the THz frequency range are drastically different from each other. We have shown that $\sigma_1$ is maximum and $\varepsilon_1$ is minimum when the $Fe^{+2}/Fe^{+3}$ ratio is equal to 0.54 which is the ratio of $Fe^{+2}/Fe^{+3}$ for pure $Fe_3O_4$. The $\sigma_1$ reduces and $\varepsilon_1$ increases at all temperatures when the $Fe^{+2}/Fe^{+3}$ ratio deviates from 0.54. We have shown that a slight change in the $Fe^{+2}/Fe^{+3}$ ratio can induce large changes in the optical properties which shall have implications in the application of the $Fe_3O_4$ films in THz spintronics.

# 1. Introduction

The $Fe_3O_4$ has been a material of interest for several decades because of its interesting electrical and magnetic properties like high Curie temperature, high saturation magnetization, 100 % spin polarization, occurrence of a Verwey transition and a debate over the exact nature of the room temperature state [1-5]. The Verwey transition in $Fe_3O_4$ is also associated with the lattice symmetry change from a high temperature cubic and metallic phase to a low temperature monoclinic and insulating phase along with a simultaneous transition in the spin state [1, 3, 5-6]. Additionally, in the recent years, there is a growing interest on the studies of $Fe_3O_4$ films deposited on semiconducting substrates due to their technological potential in spintronics and magnetic tunnel junctions [7-8]. In the spin based electronics, the half metallic $Fe_3O_4$ films [8-10] deposited on semiconducting substrate is considered a potential spin injector material for generating spin polarized currents in semiconductors [11,12]. As the feature size of the semiconducting circuits and devices goes to ~100 nm, the operational frequencies go to the higher GHz to THz range and the dielectric properties of these structures in this frequency range become important [13]. However, to the best of our knowledge, there is no report on the complex optical constants of the $Fe_3O_4$ thin films deposited on Si substrate ($Fe_3O_4$/Si) in the THz frequency range.

The properties of $Fe_3O_4$ thin films depend on the substrate chosen for deposition even in the case of highly oriented film-growth [14]. The room temperature DC electrical resistivity of the Fe$_3$O$_4$ thin films deposited on the Si substrate (Fe$_3$O$_4$/Si) is (1 mΩ-cm) an order of magnitude lower than that (0.01 Ω-cm) of the Fe$_3$O$_4$ films deposited on sapphire substrate (Fe$_3$O$_4$/Al$_2$O$_3$). The change in the electrical resistivity across the Verwey transition ($T_V$ ~120 K) is sharp and abrupt in the Fe$_3$O$_4$/MgO films (similar to bulk single crystals) whereas it is a broad and smooth changeover in the Fe$_3$O$_4$/Si films [8, 14]. A few reports are available in the literature on the optical properties of bulk Fe$_3$O$_4$ over a wide frequency ($\omega$) and temperature ($T$) range [3, 15-20]. Pimenov et. al. have reported that the dynamical conductivity $\sigma_1(\omega)$ of bulk single crystal at 295 K is about 250 Ω$^{-1}$cm$^{-1}$ and is independent of $\omega$ in the THz range [18]. However, the room temperature value of $\sigma_1(\omega)$ in the GHz-THz range reported by various authors vary by orders of magnitude and the reason behind this is not yet clear [15-16]. Even the $\omega$ dependence of $\sigma_1$ changes when the thickness of the film is reduced to the nanometer scale [21]. Therefore, from the point of view of technological applications, it is important to understand the optical properties of the Fe$_3$O$_4$ thin films, especially in the GHz-THz range.

Here we have shown that for the Fe$_3$O$_4$/Si films, the $\varepsilon_1(\omega, T)$ is minimum and $\sigma_1(\omega, T)$ is maximum for the Fe$^{+2}$/Fe$^{+3}$ ratio of 0.54 which is the ratio of Fe$^{+2}$/Fe$^{+3}$ for pure Fe$_3$O$_4$. We have also shown that the both $\varepsilon_1(\omega, T)$ and $\sigma_1(\omega, T)$ changes drastically with a slight deviation of Fe$^{+2}$/Fe$^{+3}$ ratio from 0.54.

## 2. Experimental Details

Four thin films of Fe$_3$O$_4$ were deposited on Si (100) substrate using pulsed laser deposition technique in the standardized protocol [14, 22]. A KrF excimer laser source ($\lambda$ = 248 nm, pulse width = 20 ns) was used to ablate the α-Fe$_2$O$_3$ target. The pulse repetition rate was set at 10 Hz and the energy density of the laser beam at the target was 2 J/cm$^2$. The deposition was performed at the substrate temperature of 450 $^O$C, under a background pressure of ~10$^{-6}$ Torr. The target-to-substrate distance was 5 cm. After deposition, the sample was cooled at the rate of 2 $^O$C/ min in the same environment which was used during the deposition. The structural characterization of the samples was done with the help of X-ray diffraction (XRD) using Cu-Kα radiation in Bragg-Brentano configuration. The XANES measurements were performed on the scanning EXAFS beam-line (BL-9) of the Indus-II synchrotron at RRCAT, Indore. Atomic force microscopy (AFM) was performed on a Bruker (MultiMode 8-HR) machine.

The $T$ dependence of the DC electrical resistivity ($\rho$) of the $Fe_3O_4$/Si films were measured in a magneto-optical cryostat system (Cryofree Spectromag CFSM7T-1.5, Oxford Instruments, UK). The $T$ dependence of magnetization ($M$) was performed in a superconducting quantum interference devised based vibrating sample magnetometer (SQUID-VSM, MPMS 3, Quantum Design, USA). The $M(T)$ measurements were performed following the zero field cooled warming (ZFC) protocol, where the sample was cooled down from the room temperature to 2 K in zero magnetic field. A field of 100 Oe was then applied at 2 K, and the measurements were performed while warming-up the sample.

The dielectric constants of the $Fe_3O_4$/Si films were measured in the range $T$ = 2-300 K and $\omega/2\pi$ = 0.2- 1.2 THz using a custom built THz time-domain spectrometer (Teravil Ltd./ Ekspla uab, Lithuania) designed around the cryogen free magneto-optical cryostat system that was also used for the measurement of $\rho$ mentioned above. A piece of the Si (100) wafer that was used as a substrate for the deposition of the $Fe_3O_4$ films was used as the reference for the THz time domain measurements. The electrical amplitude of the transmitted THz radiation through the $Fe_3O_4$/Si film as well as the reference were recorded one after the other at each temperature and an averaging was done over 500 signal-samples to increase the signal to noise ratio.

## 3. Results and Discussion

The XANES measurements performed in the fluorescence mode were used to find the ratio of the Fe ions ($Fe^{+2}$/ $Fe^{+3}$ ) in the present $Fe_3O_4$/Si films. These measurements are sensitive up to micron level thickness and hence the obtained information is related to the bulk of the films. The ratio ($Fe^{+2}$/ $Fe^{+3}$ ) was found to be 0.22, 0.39, 0.54 and 0.75 respectively for the films with 20 nm (Film-A), 115 nm (Film-B), 200 nm (Film-C) and 200 nm (Film-D) thickness. It may be noted that for pure $Fe_3O_4$ powder the ratio of $Fe^{+2}/Fe^{+3}$ is found to be 0.54 which indicates the presence of secondary phases in films A, B and D.

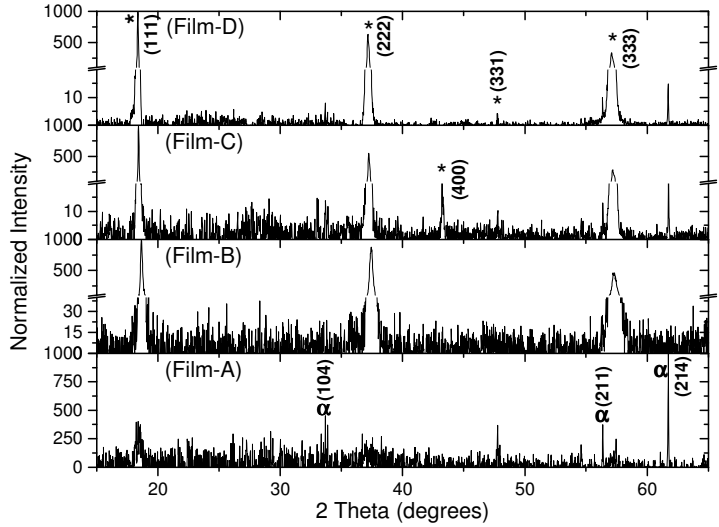

Fig. 1: XRD pattern for the $Fe_3O_4$/Si films. The peak marked with a * represents the $Fe_3O_4$ phase and that with an $\alpha$ represents the $\alpha$-$Fe_2O_3$ phase. The XRD pattern for Film-A has the major contribution coming from the $\alpha$-$Fe_2O_3$ phase. While for rest of the films the major contribution is from the $Fe_3O_4$ phase.

Figure 1 shows the XRD patterns for the different $Fe_3O_4$ films. Data above 65 degrees is not shown in order to avoid the high intensity Si reflections for the sake of clarity. All the films are observed to have grown with the preferred orientation [111] with cubic structure. The contribution from the second phase, i.e. $\alpha$-$Fe_2O_3$, has also been observed in all the films. The peak positions for the respective phases are marked in the patterns. The Film-A has the major contribution of 60 % from the $\alpha$-$Fe_2O_3$ phase, and this is found to consistent with the resistivity and magnetization measurements presented below. All the other films are predominantly $Fe_3O_4$ with contribution of < 3%, < 3% and < 2 % respectively from the $\alpha$-$Fe_2O_3$ phase for films the B, C and D. The calculated lattice parameters of the $Fe_3O_4$ phase in the films A, B, C and D are 8.403 Å, 8.385 Å, 8.372 Å and 8.381 Å respectively.

Figure 2 shows the surface morphology of the films obtained using AFM. The films B, C and D are continuous with the estimated grain sizes of 38.6 nm, 109 nm and 97 nm respectively. The Film-A (20 nm thick) shows dark voids in between grains of average size of 49 nm which indicates the discontinuous growth of this film. The Film-D shows the signatures of agglomeration of particles. The average roughness $R_{max}$ for the films A, B, C and D are 2.18 nm, 1.54 nm, 2.33 nm and 2.83 nm respectively.

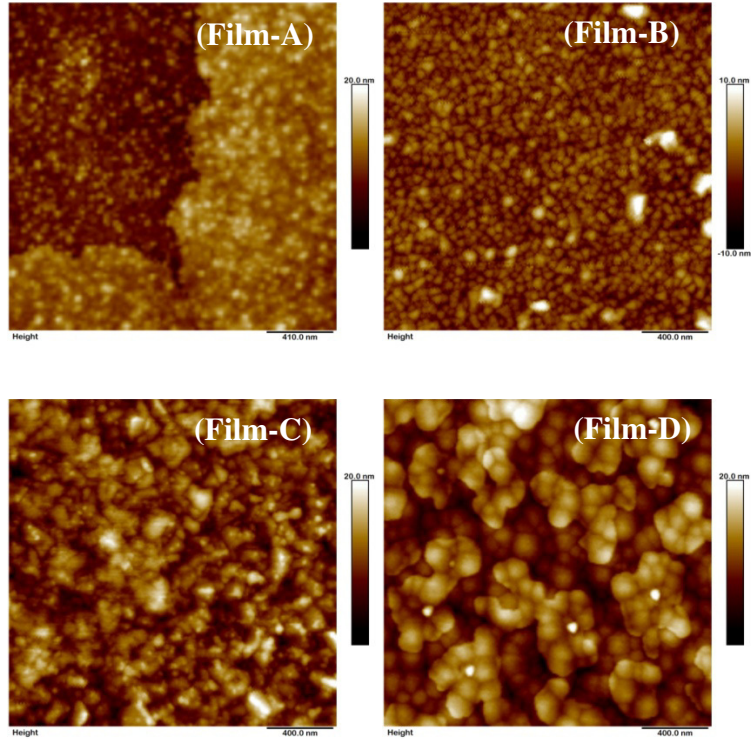

Fig. 2: AFM pictures showing a discontinuous growth for the Film-A and continuous growth for the films B, C and D. Agglomeration of particles is observed for the Film-D.

Figure 3(a) shows the $T$ dependence of $\rho$ of the $Fe_3O_4$/Si films in zero field from 90- 300 K. While $\rho$ increases with decreasing $T$ at all the measured $T$, a change of slope near $T_V$ ~121 K [1, 6, 23] is observed in the films B, C and D. This change of $\rho$ near $T_V$ is quite clearly observed in the films C and D, but is comparatively less significant in the Film-B. The change in the $\rho$ around $T_v$ in any of the present $Fe_3O_4$/Si films is not as sharp as that reported for the single crystal samples [3, 18, 24]. This may be due to the stress induced in the films during deposition [14, 25] on Si. No detectable signature of the Verwey transition is observed in the Film-A. On the other hand, a clear transition is visible in the $\rho(T)$ of Film-A at T ≈ 263 K. It is reported in literature [26] that $\alpha$-$Fe_2O_3$ exhibits a first order magnetic phase transition from canted anti-ferromagnetic to anti-ferromagnetic phase (with decreasing $T$) at ~263 K which is referred to as the Morin transition. Since 60 % $\alpha$-$Fe_2O_3$ phase was detected in the XRD pattern of Film-A, the change of $\rho$ observed near 263 K may be ascribed to the Morin transition in the $\alpha$-$Fe_2O_3$ phase in this film. It is observed that the $\rho$ of this film is an order of magnitude higher than those of the rest of the films at $T$< 263 K. The Film-B also seems to show a weak signature of this Morin transition.

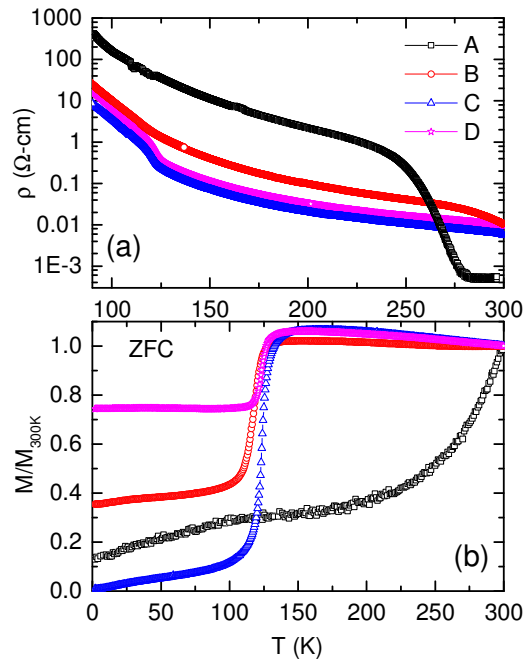

Fig. 3: Temperature dependence of (a) DC electrical resistivity of $Fe_3O_4$/Si films in zero magnetic field and (b) relative magnetization (M/$M_{300K}$) of the $Fe_3O_4$/Si films in 100 Oe field. Across the Verwey transition, a sharp change in magnetization is observed in the films B, C and D, while only a weak change of slope is observed in the Film A.

Figure 3(b) shows the $M(T)$ curves for the $Fe_3O_4$/Si films relative to the $M$ at 300

K. A clear change in $M/M_{300K}$ is observed across the $T_v$ near 119 K, 124 K and 123 K for the films B, C and D respectively (determined using the $T$ dependence of the derivative of M(T)) and these are in line with the previous studies [24, 27]. However, such a jump in $M$ is not observed in the Film-A. Only a weak change of slope in $M$ is observed at 98.5 K. On the other hand, a rapid drop in $M$ is observed as the $T$ is lowered from 300 K to 250 K. This may be related to the Morin transition discussed above.

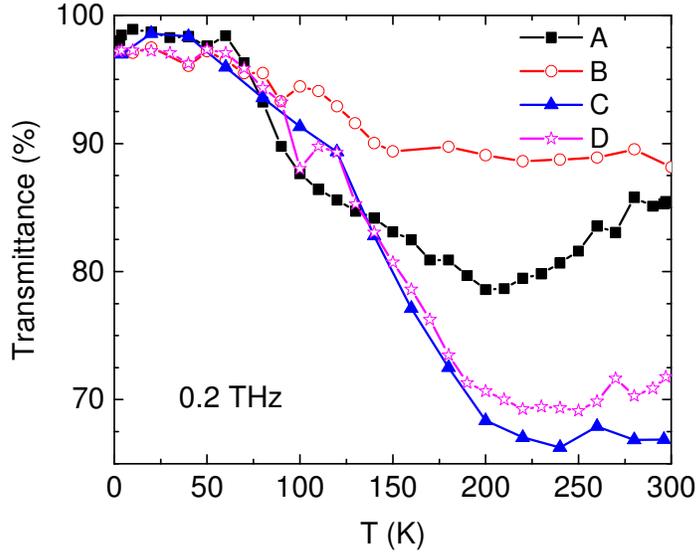

Fig. 4: Terahertz transmittance as a function of temperature at 0.2 THz for the $Fe_3O_4$/Si films. The relative transmission with respect to the substrate at room temperature for the films A, B, C and D are 85, 88, 67 and 72% respectively. At temperatures much below the Verwey transition, the transmission through the films is nearly equal to that through the substrate.

Both the amplitudes $|E_{Fe3O4/Si}|$ and $|E_{ref}|$ and the phases $\Phi_{Fe3O4/Si}$ and $\Phi_{ref}$ of the transmitted THz signal were obtained by taking the Fourier transform of the time traces of the THz pulses transmitted through the film and the reference substrate respectively. The transmittance through the $Fe_3O_4$ film was estimated by taking the ratio as Transmittance = $|E_{Fe3O4/Si}|/|E_{ref}|$. The multiple reflections (Fabry-Perot reflections) from the substrate were truncated by appropriately setting the position of the initial delay.

Figure 4 shows the $T$ dependence of transmittance through the $Fe_3O_4$ films at 0.2 THz. At room temperature, the transmittance for the films A, B, C and D are 85, 88, 67 and 72 % respectively. For all the films except Film-A, the transmittance increases with decreasing $T$ and becomes close to 100 % below 60 K which is well below the $T_v$. The changes in the THz transmittance across $T_v$ in the films C and D are substantial, while the same is quite small in Film B. For the film A, however, the THz transmittance initially decreases with decreasing $T$ below 300 K, and then increases continuously with further decrease of $T$ below 200 K. This initial decrease in transmission below 300 K may be due to the Morin transition in the $\alpha$-$Fe_2O_3$ phase of this film. The difference in the transmission between the high $T$ and low $T$ phases is observed to be most significant at the lower frequencies and becomes weaker with the increasing frequencies (not presented).

The optical constants were estimated from the amplitude and the phase of the transmitted THz using the formalism given by Kida et. al. [28]. The frequency dependence of the complex optical properties of the Si substrate for each film was ascertained independently and was subsequently used to estimate the optical constants of the respective $Fe_3O_4$ films.

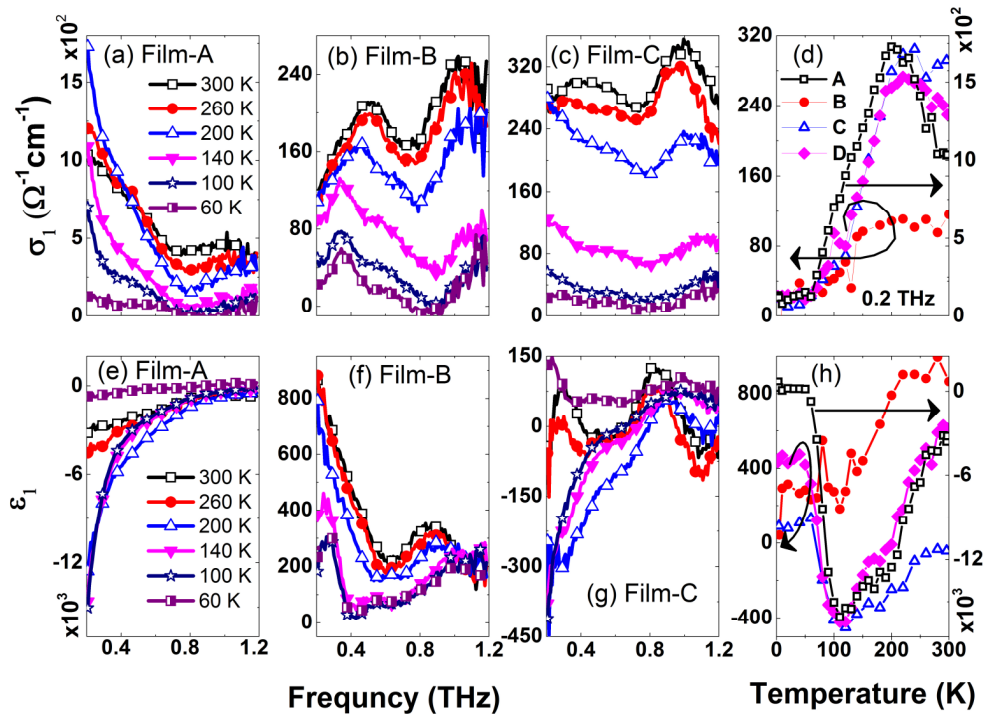

Fig. 5: Frequency dependence of $\sigma_1$ and $\varepsilon_1$ for Film-A [(a) and (e)], Film-B [(b) and (f)]

and Film-C [(c) and (g)] respectively at different temperatures. (d) and (h): Temperature dependence of $\sigma_1$ and $\varepsilon_1$ of the $Fe_3O_4$/Si films estimated from the frequency dependence at different temperatures (right axis for film-A and left axis for films B, C and D).

The figures 5(a), (b) and (c) show the real part of the optical conductivity $\sigma_1(\omega/2\pi)$ of the $Fe_3O_4$/Si films A, B and C respectively at different $T$. For these samples, the $\sigma_1(\omega/2\pi)$ is higher at the room temperature and then decreases with decreasing $T$. However the frequency dependence of $\sigma_1$ is different for the films A, B and C. The frequency dependence of Film-D is same as that of Film-C, and is not shown here separately. The room temperature values of $\sigma_1$ at low frequencies are found to be consistent with the DC conductivity obtained from the measurement of $\rho$. For the Film-A, contrary to the higher $\rho$ measured at lower temperatures, the optical conductivity is found to be much higher than the other films. It may be related to the oxygen stoichiometry in the film, since it is known that that the oxygen deficiencies leads to the increase in optical conductivity from $10^{-6}$ to $10^{-3}$ Ohm$^{-1}$cm$^{-1}$ in the ferrite samples [29].

For better understanding of the optical properties across the Verwey transition, we have plotted in figure 5(d) the $T$ dependence of $\sigma_1$ at 0.2 THz for all the four samples. For the Film-A, the $\sigma_1(T)$ initially increases when $T$ reduces from 300 K down to 250 K. When $T$ is reduced below 250 K the conductivity decreases further. In view of the signatures obtained in the XRD pattern, and the $T$ dependence of $\rho$ and $M$ presented above, this peak in $\sigma_1(T)$ at 250 K may be associated with the Morin transition observed in the $\alpha$-$Fe_2O_3$ phase of Film-A. Subsequently, $\sigma_1(T)$ shows a drop across the Verwey transition, which becomes less sharp with increasing frequency (not shown here). It may be noted that the drop in conductivity occurs over a large $T$ range of 240-60 K which is in line with the increase of THz transmittance across the Verwey transition with decreasing $T$. This indicates that the low $T$ insulating and the high $T$ metallic/semiconducting phases coexist in the $T$ range of 240-60 K. The Verwey transition is reported to be of first order nature [18, 23], and the nucleation, growth and coexistence of phases are expected across such a phase transition [30, 31]. Though the ideal first order phase transition is expected to be sharp, it may be broadened [31] due to the influence of disorder [32] and strain [33].

Thus the previous observation of the '*precursors for the metal to insulator transition in the Fe$_3$O$_4$/Si film*' starting well above the $T_V$ [25] is consistent with the indication of a very broad phase coexistence regime across the first order phase transition. This again indicates towards a major influence of the strain produced in the Fe$_3$O$_4$/Si films on the Verway transition, and the possibility of strain-disorder coupling across the transition [33].

The figures 5(e), (f) and (g) show the $\varepsilon_1(\omega/2\pi)$ at different $T$ for the films A, B and C respectively. The $\varepsilon_1(\omega/2\pi)$ is observed to be very different for different films. For the Film-A, the $\varepsilon_1(\omega/2\pi)$ is negative at all $T$. This is consistent with the higher optical conductivity of this film, as a negative $\varepsilon_1$ is generally observed for the free carriers [34]. For the Film-B, the $\varepsilon_1(\omega/2\pi)$ is positive at all $T$. For the Film-C, the $\varepsilon_1(\omega/2\pi)$ is negative at room temperature. For $T < 260$ K the $\varepsilon_1(\omega/2\pi)$ rapidly drops to larger negative values with decreasing frequency. In the insulating state for $T \ll T_V$, the $\varepsilon_1(\omega/2\pi)$ becomes positive and increases with increasing frequency. As $T$ decreases, the $\varepsilon_1(T)$ at different frequencies for the Fe$_3$O$_4$/Si films A, C and D [figure 5(h)] drops sharply to the negative values and then start increasing below the Verwey transition. For the Film-B, although $\varepsilon_1$ is positive over the entire range of temperature of measurement, we observed large change of $\varepsilon_1$ across $T_V$. The large change in the dielectric constant of all the films across the Verwey transition is associated with the dielectric catastrophe across the metal insulator transition [18, 35].

An important feature in the optical response of the present Fe$_3$O$_4$/Si(100) films is that over the entire frequency range 0.2-1.2 THz and the temperature range 2- 294 K, the magnitudes of $\varepsilon_1$ and $\sigma_1$ of the film are much larger than those for the bulk single crystals [18]. In general, this may indicate the presence of a large amount of free carriers in the Fe$_3$O$_4$/Si(100) films. Such free carriers may be present in the Fe$_3$O$_4$/Si(100) films due to the anti-phase boundaries [27, 32]. The substrate induced strain on the films may also lead to the enhancement of $\varepsilon_1$ and $\sigma_1$ in the Fe$_3$O$_4$/Si films. Recently, Grenier et al., have shown that the low $T$ monoclinic phase of Fe$_3$O$_4$ may be observed even at $T_V + 80$ K in the ultra-thin films of Fe$_3$O$_4$ on Ag substrate [25]. They have argued that the structural transition is initiated locally at such high $T$ in the thin films due to the strain resulting from the lattice mismatch between the film and the substrate [25].

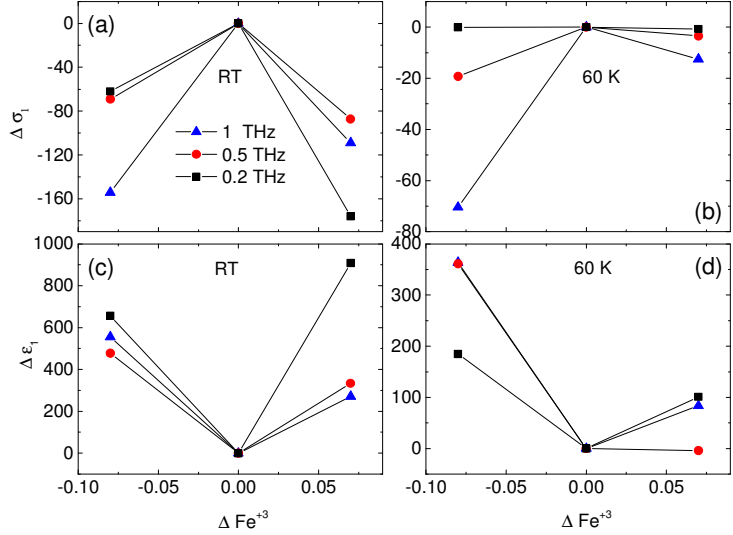

Fig. 6: Comparison of changes in the optical properties as a function of deviation from the amount of $Fe^{+3}$ present in the pure $Fe_3O_4$.

In order to understand the variation observed in the optical properties of our $Fe_3O_4$/Si films, we plot in Fig. 6, the changes in the optical properties as a function of deviation ($\Delta Fe^{3+}$) of the amount of $Fe^{+3}$ present in the film from the amount of $Fe^{+3}$ present in the pure $Fe_3O_4$. Here we have not considered the 20 nm film in this comparison as it is predominantly $Fe_2O_3$. The figures 6(a) and (b) show the change in the optical conductivity $\Delta\sigma_1(\omega, RT)$ ($T > T_v$) and $\Delta\sigma_1(\omega, 60\ K)$ ($T < T_v$) respectively. At all temperatures, the $\sigma_1(\omega, T)$ is maximum for the $Fe^{+3}$ amount of 0.65 (i.e., $\Delta Fe^{3+} = 0$). It is observed that any deviation in the amount of $Fe^{+3}$ results in the reduction of the $\sigma_1(\omega, T)$. On the other hand, $\varepsilon_1(\omega, T)$ is minimum at all temperatures for the $Fe^{+3}$ amount of $\Delta Fe^{3+} = 0$ (Fig. 6(c) and 6(d)) and increases for any deviation from this $Fe^{+3}$ amount. This indicates that growing the $Fe_3O_4$/Si films with an $Fe^{+3}$ amount as close as to 0.65 (the amount of $Fe^{+3}$ in pure $Fe_3O_4$) is important for using these materials for technological applications. Note that the amount of $Fe^{+3}$ in a film depends on many factors such as the grain size or grain boundary density, presence of secondary phases, strain, etc. Therefore, it is important to standardize the growth of $Fe_3O_4$/Si films with respect to optical/THz properties before considering these films for technological applications.

## 4. Summary and Conclusion

We have measured the frequency dependence of the dielectric constant and the optical conductivity of the $Fe_3O_4$ films of different thicknesses deposited on Si substrate ($Fe_3O_4$/Si) in the THz range (0.2-1.2 THz) at temperatures 2-300 K. We have found that the $\varepsilon_1$ and $\sigma_1$ of the $Fe_3O_4$/Si films of different thickness vary appreciably although a clear signature of the Verwey transition is observed in these films at identical temperatures. The $\varepsilon_1$ and $\sigma_1$ of these films in the THz range are also found to be much higher than those of bulk single crystals. The difference in optical properties between the $Fe_3O_4$/Si films and the bulk seem to be related to the presence of a large number of free carriers in the films, which may arise due to the anti-phase boundaries. Substrate induced strain on the films may also play a role in the enhancement of $\varepsilon_1$ and $\sigma_1$ in the $Fe_3O_4$/Si films. We have observed that for the $Fe_3O_4$/Si films, the $\varepsilon_1(\omega, T)$ is minimum and $\sigma_1(\omega, T)$ is maximum for the $Fe^{+2}/Fe^{+3}$ ratio of 0.54 which is the ratio of $Fe^{+2}/Fe^{+3}$ for pure $Fe_3O_4$. We have shown that both $\varepsilon_1(\omega, T)$ and $\sigma_1(\omega, T)$ change drastically with a slight deviation of $Fe^{+2}/Fe^{+3}$ ratio from 0.54. Our studies suggest that the $Fe^{+2}/Fe^{+3}$ ratio, morphology and the grain size varies significantly between the films of different thickness even when the films are deposited in identical growth conditions. This leads to drastic difference in the optical properties of the films in the THz range. Thus, in order to produce $Fe_3O_4$ films having reproducible properties at THz frequencies, which will be a requirement for the technological applications, there is a need to fine tune the deposition parameters so that the $Fe^{+2}/Fe^{+3}$ ratios may be controlled and reproducible morphology and grain sizes can be obtained.

## Acknowledgments

We thank Dr. S. B. Roy for his interest in this work. We also thank Dr. S. N. Jha and Dr. Ashok Kumar Yadav for the XANES measurements.